\documentclass[sigconf]{acmart}

\usepackage{booktabs} 

\usepackage{amssymb}

\acmDOI{XXX}

\acmISBN{XXXX}

\acmConference[IFUP 2018]{IFUP 2018}{Feb 2018}{LA} 
\acmYear{2018}
\copyrightyear{2018}

\begin{document}
\title{Fusing Multifaceted Transaction Data for User Modeling and Demographic Prediction}

\author{Yehezkel S. Resheff}
\affiliation{%
  \institution{Intuit}
}
\email{Hezi_Resheff@intuit.com}

\author{Moni Shahar}
\orcid{1234-5678-9012}
\affiliation{%
  \institution{Intuit}
}
\email{Shimon_Shahar@intuit.com}



\begin{abstract}
Inferring user characteristics such as demographic attributes is of the utmost importance in many user-centric applications. Demographic data is an enabler of personalization, identity security, and other applications. Despite that, this data is sensitive and often hard to obtain. Previous work has shown that purchase history can be used for multi-task prediction of many demographic fields such as gender and marital status. Here we present an embedding based method to integrate multifaceted sequences of transaction data, together with auxiliary relational tables, for better user modeling and demographic prediction.   
\end{abstract}

%
%



\maketitle

\section{Introduction}
Demographic characteristics of an individual such as age, gender, and marital status are used in many real-world applications. For example, when designing a marketing campaign one of the first decisions taken is the target population, this population is defined among other things by demographic characteristics \cite{jansen2010gender}. Content and user experience in many applications is modulated by user characteristics, aiming to provide the best possible version to each individual. Another family of applications for which demography is crucial is insurance; for instance, demographic characteristics are required for life expectancy estimation, and specific demographic variables such as gender and marital status are also relevant for car insurance, house insurance, and others. 

However, these two examples are fundamentally different. When filling an insurance form one must provide all the required information, or else the application will be rejected. On the other hand, marketing and advertisement professionals often have partial demographic data at best. Although in many occasions people are asked to provide their personal information, the provider may have no effective way to enforce disclosure of full and accurate demographic details. Furthermore, many users are reluctant to give this data due to privacy concerns. Another complication with demographic data is how to keep it up to date. Many life events change demographic status, so even if when acquired the data was accurate, this may no longer be the case. Demanding that all users  keep their information up to date is possible only for mandatory actions like filling a government form.  

The above discussion suggests that inferring demographic from available data is of great value. Specifically, for financial applications with access to much of a user's financial data, it is important to be able to model the user and infer demographic attributes based on it. Our first and foremost motivating applications is to personalize and deliver more relevant content in a better and user-suited experience. In addition, we discuss the application of the proposed method to detecting change in a user's profile triggered by life-time events and the related task of stolen identity prevention. 

In the current applications the data available in real time is bank transaction data. For some users we also have partial demographic variables available as well as other data. A minority of the users provide full demographic information. In this work we focus on predicting demographics based solely on the sequences of transactions, augmented by relational information. This data is guaranteed to exist for all active users. The main contribution of this paper is in presenting a novel method for fusing multiple sequences of categorical information, together with structured relational data, for an end-to-end multi-task prediction of demographic characteristics.

\section{Related Work}

User modeling for demographic prediction has long been a topic of interest both in academia and in industry. Methods have been proposed for inferring demographics based on browsing behavior and search queries \cite{hu2007demographic,bi2013inferring,culotta2015predicting}, social links \cite{zhuang2006demographic,dong2014inferring}, and mobile behavior \cite{ying2012demographic,malmi2016you}. These works, as well as other older works, apply hand crafted features for the predictive modeling.  

Our method draws inspiration from a previous line of work \cite{wang2016your}, in which a method is proposed to learn a user representation for multi-target prediction of demographic attributes, based on a single sequence of super-market purchases. Our method can be seen as a generalization to multiple sequences, combined with auxiliary relational information and a deep representation component.  

\section{Method}
\label{sec:method}

\begin{figure*}
\centering{}
\includegraphics[trim={0 0 2.5cm 1.5cm},clip,width=.95\linewidth]{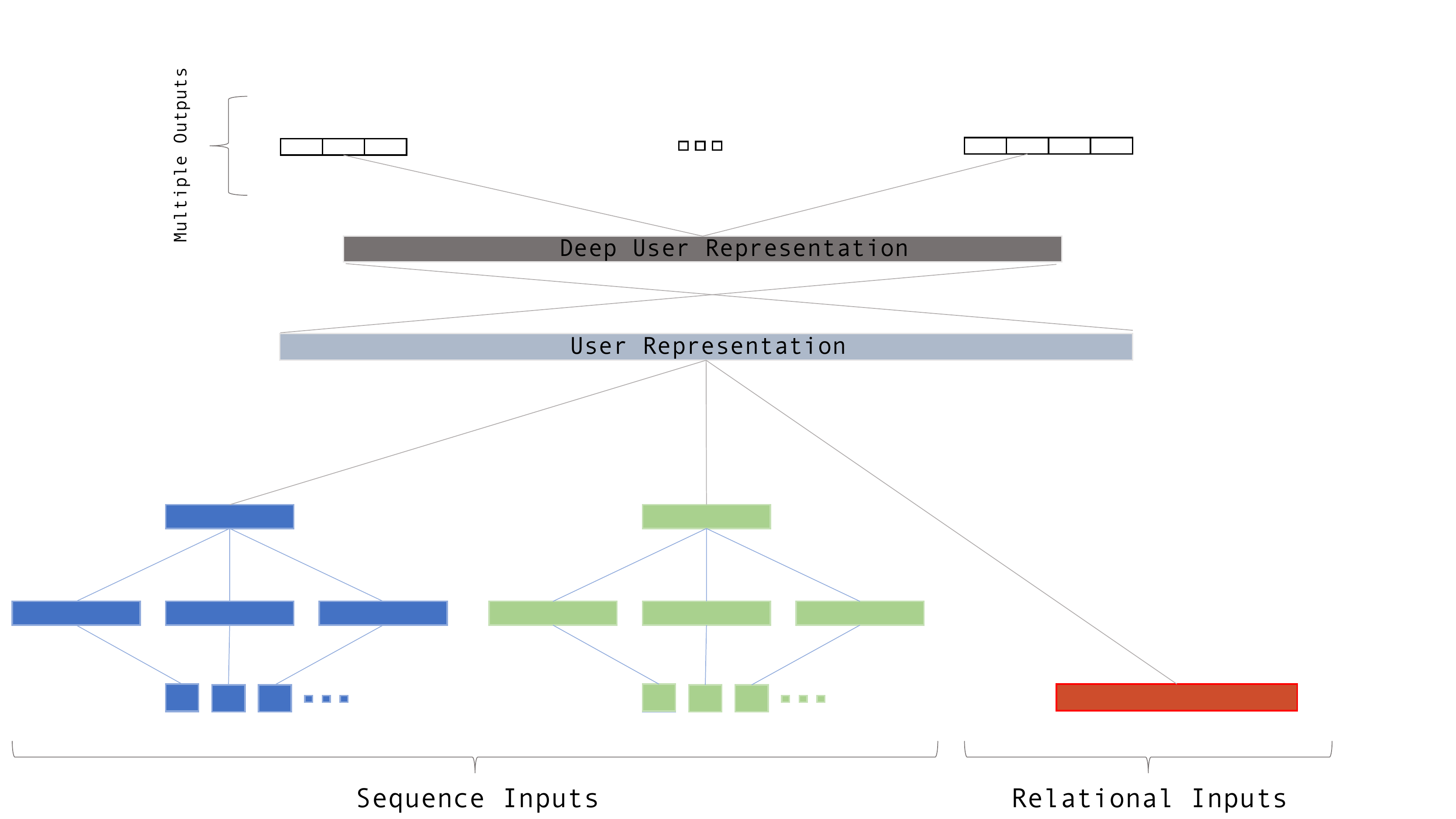}
\caption{The overall structure of the model. }
\label{fig:model-graphic}
\end{figure*}

We formulate the demographics prediction problem in terms of multi-task learning based on heterogeneous inputs. Inputs take two shapes: sequences of categorical variables, and numerical information. Outputs are all assumed to be categorical. Formally, we are interested in learning functions of the form:

\begin{equation}
f: S_1^{<\infty} \times \dotsc \times S_n^{<\infty} \times \mathbb{R}^p \rightarrow Y_1 \times \dotsc \times Y_m
\end{equation}

\noindent where $S_i$ are finite sets representing categorical information (such as the identity of the merchant a purchase was made from), and likewise finite sets $Y_j$ are  categorical demographic information to be predicted (such as education). The $S^{<\infty}$ notation is used to denote a finite sequence of elements of set $S$. 

We adopt an embedding approach, where we represent each of the variable length sequences as a constant size vector. This is done using an aggregation function (such as averaging) over the  embeddings vectors representing the individual elements (see the blue unit at the bottom left of Figure \ref{fig:model-graphic}). Much like aggregating word vectors in NLP tasks, this approach transforms the unstructured sequence problem into a constant size vector representation amenable to further processing. 

For each of the input sequences $\{s_j\}: s_j \in S_i$, we define an element embedding function:

\begin{equation}
g_i: S_i \rightarrow \mathbb{R}^{l_i}
\end{equation}

\noindent and the sequence $\{s_j\}$ of elements of $S_i$ is then represented as:

\begin{equation}
\textrm{embedding}_i(\{s_j\}) = \frac{1}{|\{s_j\}|} \sum_j g_i (s_j)
\end{equation}

In practice, since we assume nothing about the embedding functions, and would rather rely on the data and classification task in order to learn a meaningful representation of the sequence elements, the $g_i$ functions are represented non-parametrically as matrices $G_i$. The $j-th$ column of $G_i$ is thus the embedding of the $j-th$ element of the set $S_i$. 

Following the embedding of the sequences, the derived representations are concatenated together with the auxiliary numerical information to form the raw user representations. The final deep user representation is  obtained by feeding the raw representation into a fully-connected neural network. All $m$ demographic fields are then inferred from this layer using $m$ softmax classification layers (See figure \ref{fig:model-graphic} for a description of the full end-to-end architecture).

The full model is then trained on the combined (categorical cross-entropy) losses related to each of the individual target. This combined end-to-end approach allows the learned user representation to combine and share information from all the prediction tasks, towards a unified view of the a user as a whole. 

We further propose that fine-tuning the model with respect to each of the targets could benefit the individual target performance. In the fine-tuning experiments (section \ref{sec:results}) the model is first trained in the multi-task setting described above. Then, all but a single softmax output are removed together with the associated weights, and the partial network is fine-tuned using the categorical cross-entropy loss of the remaining target.

\section{Experiments}
\label{sec:results}

\begin{table*}
  \caption{Comparison of the proposed model (Figure \ref{fig:model-graphic}) to alternative baselines. See section \ref{subse:baselines} for a full discussion. }
\label{tbl:results}
\begin{center}

\begin{tabular}{lcccccc}
\toprule
{} & gender & marriage status & household adults & household children & education level & residential status \\
\midrule
baseline:arg-max                &    68.0 &           45.5 &             57.6 &               50.1 &            68.6 &               42.8 \\
baeline:stacking       &    82.1 &           72.0 &             63.9 &               62.3 &            67.0 &               70.1 \\
baseline:PCA           &    80.0 &           70.5 &             63.1 &               59.5 &            67.4 &               69.5 \\
model                  &    \textbf{86.2} &           \textbf{75.4} &             \textbf{66.7} &               \textbf{65.8} &            \textbf{69.2} &               \textbf{73.2} \\
mdoel-finetuned        &    \textbf{86.2} &           75.3 &             66.5 &               \textbf{65.8} &            69.1 &               73.1 \\
\bottomrule
\end{tabular}

\end{center}
\end{table*}

In this section we present experiments comparing our method to standard baselines, and demonstrating the ability of our proposed method to model users for the purpose of demographics prediction. We further investigate the effect of model attributes such as embedding size and network depth, and test the idea of fine-tuning with respect to individual targets. 

\begin{figure}
\centering{}
\includegraphics[width=\columnwidth]{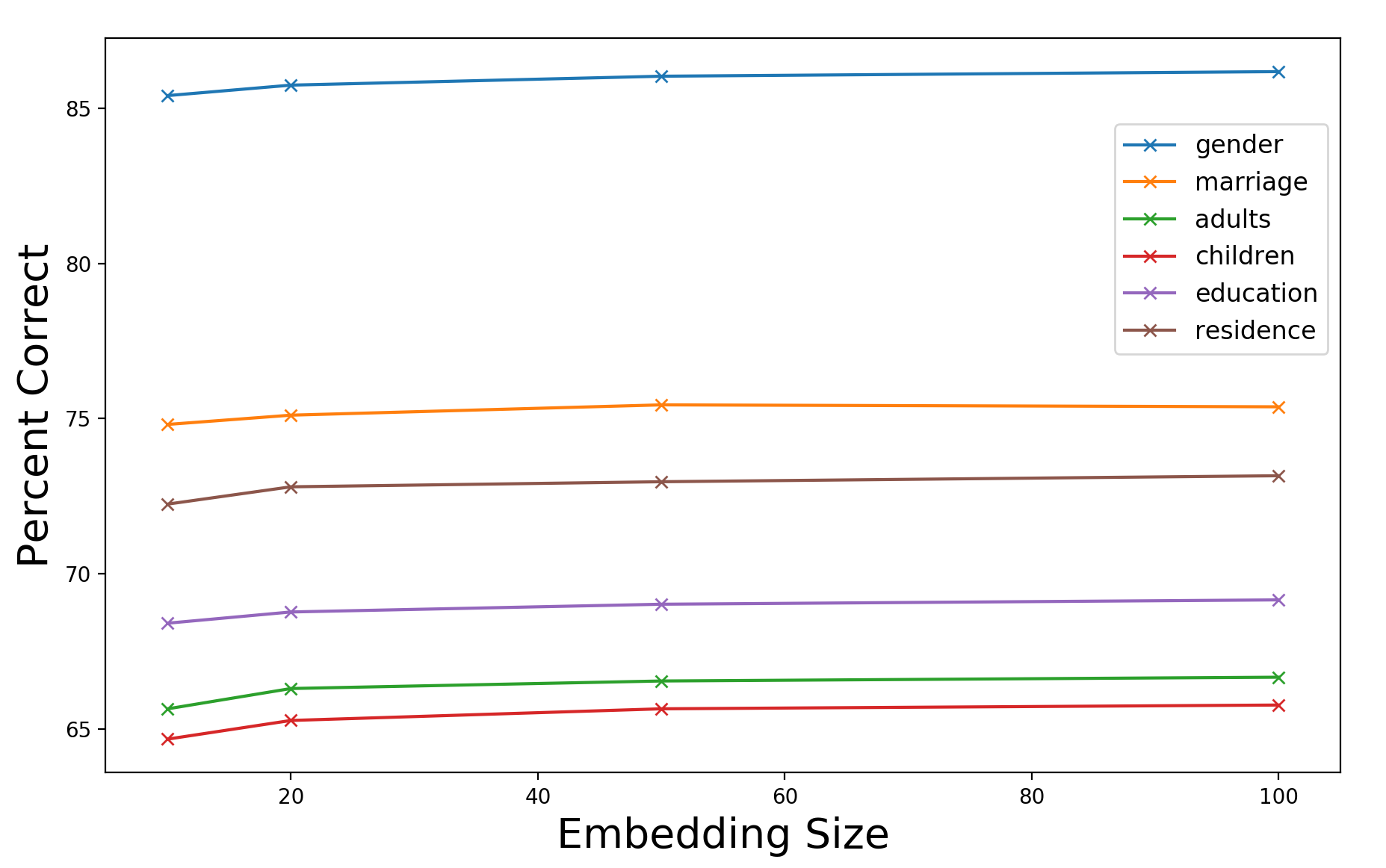}
\caption{Model accuracy (no fine-tuning) as a function of the embedding size (of the category and merchant sequences), for each of the demographics attributes. Results indicate that a size $50-100$ embedding of sequence elements is sufficient.}
\label{fig:emb-size}
\end{figure}

\subsection{Experimental Setup}
\subsubsection{Data}
Experiments are conducted on real-world user data collected on a large scale by a US based financial management software product. A total of $768,930$ users were sampled based on recent activity, and availability of at least partial demographic information. Of these, the $6$ target demographics fields (gender, marital status, number of adults in the household, number of children in the household, education level, and residential status) were all available for $682,799$ users. These were selected as the sample for our experiments in the current study.

The features used for demographic modeling are of two types. The first type is sequence information related to bank transactions and includes the category of purchases, as well as the merchant names. The second type of information is structured data on the type and number of financial accounts held by the user. Moreover, for part of the users partial demographic information is available as well as other relevant structured data. However, for the purpose of this current work, to investigate the extent these data contain the demographic information, we ignored this partial data restricting ourselves to focus on the worst case scenario necessitating full demographic profile prediction. 

\begin{figure}
\centering{}
\includegraphics[width=\columnwidth]{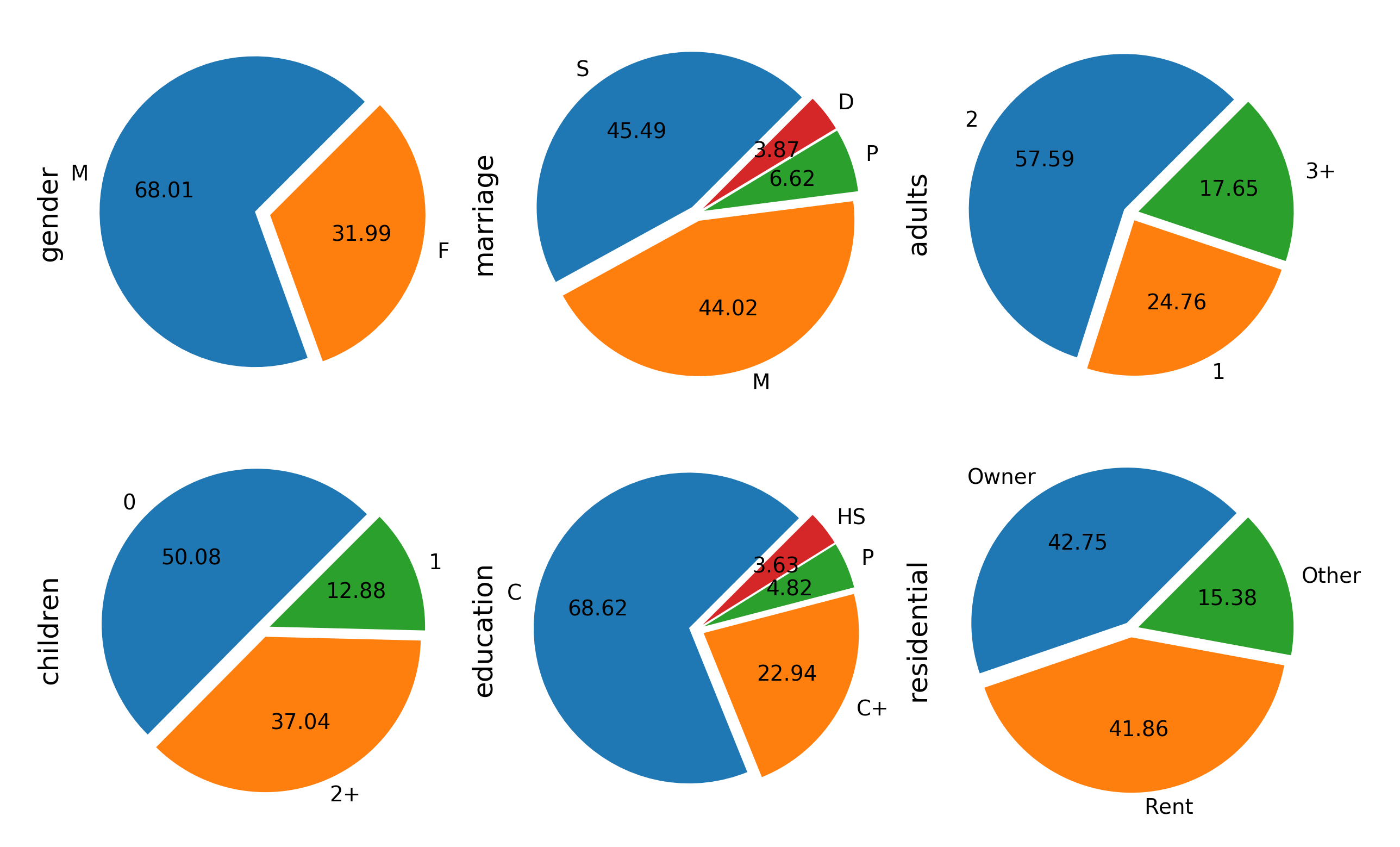}
\caption{Distributions of demographic variables in the data. Abbreviations: gender -- M (male), F (female); marital status -- S (single), D (divorced), P (partner), M (married); Education -- HS (high-school), P (professional), C (college), C+ (post-grad)}
\label{fig:demographics}
\end{figure}

The sample population is heavily male skewed with $68.01\%$ vs. $31.99\%$ female (Figure \ref{fig:demographics}). Even more notable is the apparent lack of the divorced marital status (under $4\%$). These and other limitations are inherent to this sort of user data, and limit (at least to a certain extent) the ability to use these provided labels for supervised learning in a straight forward way. 

Further complication often results from self-selection bias in the identity of users who provide demographic information. This is not so much of an issue for the current data, because the existence or absence of demographic records is related to version of the product used, rather than choice of the users themselves.  

\subsubsection{Methods and Baselines}
\label{subse:baselines}

We compare the proposed method to $3$ baselines:

\begin{enumerate}
\item arg-max: for each target, always predict the category with the highest proportion. This is a sanity-check baseline producing the accuracy available to a classifier with no feature information. 

\item stacking: raw features are stacked horizontally and a logistic regression is applied. Sequences are represented as a distribution over the elements they are comprised of.

\item PCA per sequence type: for each sequence, PCA is applied to the user-category proportion matrix, and the first 50 components are kept. These are stacked horizontally together with the relational input data and a logistic regression is applied as before. This baseline represents a simple linear approach to sequence embedding.   
\end{enumerate}

Two variants of the proposed method are tested:

\begin{enumerate}
\item the multi-task setting: the end-to-end multi-input multi-output model is trained as described in section \ref{sec:method}, and in Figure \ref{fig:model-graphic}. This models contains $314,219$ parameters. 

\item fine-tuning: the full model is trained as before. Then, for each output, the sub-network pertaining to that classification task is fine-tuned (i.e. all other outputs are discarded, producing effectively a model with a single output layer). 
\end{enumerate}

In both cases, all models are trained for 500 epochs, with an early stopping criterion. In practice, convergence is achieved after a few dozen epochs at most.

\subsection{Results}

First, we compare the proposed method to the aforementioned baselines for demographic information prediction (results are summarized in Table \ref{tbl:results}). Overall, the proposed method outperforms the baselines in all demographic fields. The margin varies substantially, from several percent in gender and marital status prediction to under one percent for education level. 

Somewhat surprisingly, model fine-tuning with respect to each of the individual targets does not improve upon the multi-target model results. Furthermore, for $4$ out of the $6$ targets, fine-tuning leads to slightly diminished results. We suggest that the multi-target model objective may provide some additional regularization, and hence reduces over-fitting, an effect which is then destroyed by the fine-tuning with a single target. 

Next, we turn to the effect of some of the major design parameters of the proposed model. With respect to the embedding size of sequences elements (Figure \ref{fig:emb-size}), the major improvement in classification performance in all the demographics prediction tasks is obtained when increasing from $10$ to $50$. Minor additional improvement is obtained with size $100$ embeddings. Result in Table \ref{tbl:results} are given for a model with an embedding size of $100$ for each of the two sequences.

With respect to the depth of the fully connected sub-network moving from user representations to deep user representations (see Figure \ref{fig:model-graphic}), only a minor improvement is achieved when increasing from $0$ to $1-2$ layers. This implies that the majority of the representational work is done by the sequence embedding mechanism, and further processing of this initial user representation is redundant to a large extent. We note that using a larger amount of structured information (red bar in Figure \ref{fig:model-graphic}) could further benefit the deep user representation. Results in Table \ref{tbl:results} are giving for a model with a depth of $1$ in the fully connected part.

\begin{figure}
\centering{}
\includegraphics[width=\columnwidth]{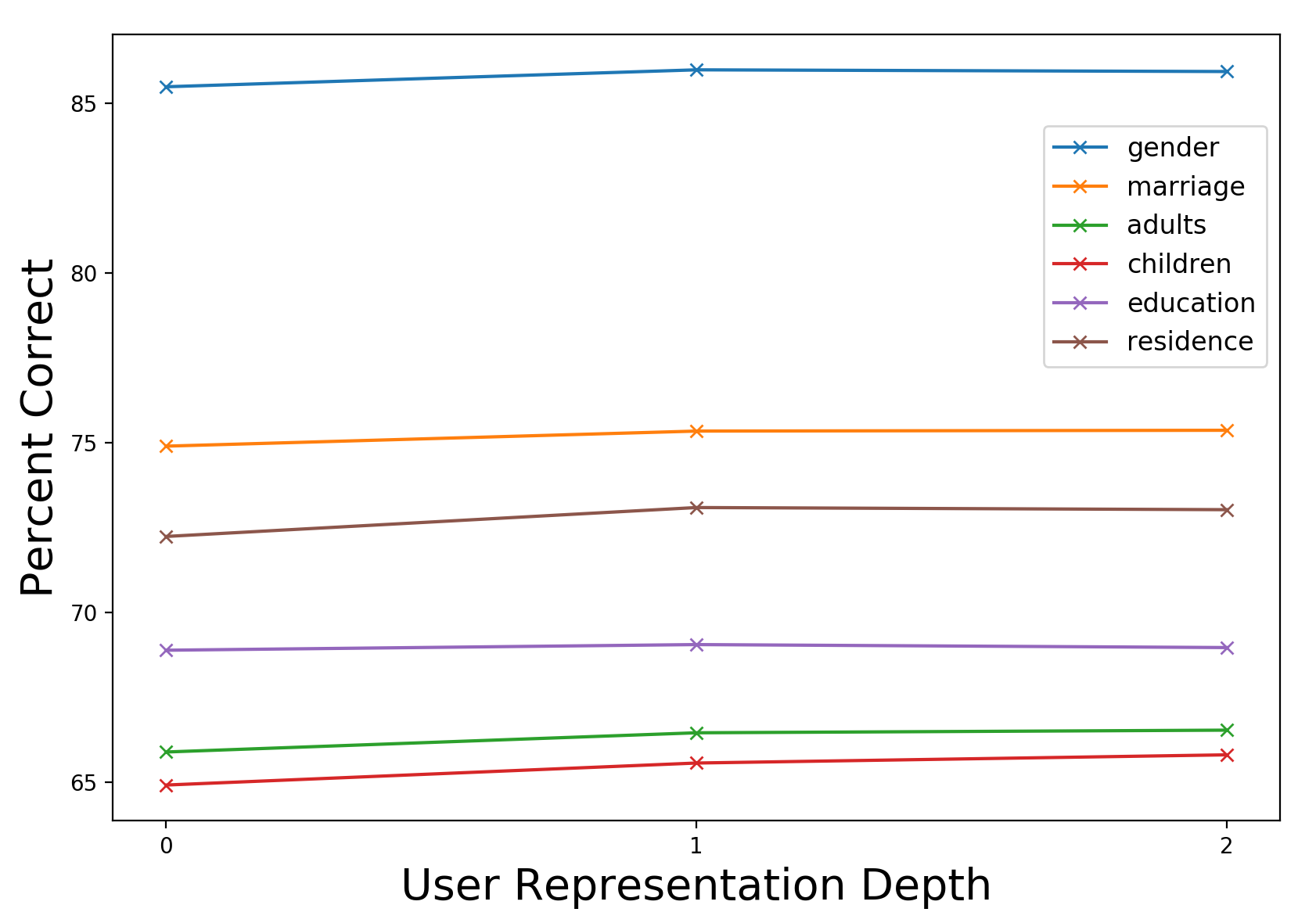}
\caption{Model accuracy (no fine-tuning) as a function of the depth of the user embedding to deep user embedding fully connected sub-network. Sequence embedding size was set to $50$. Results indicate that $1-2$ layers are sufficient.}
\label{fig:depth}
\end{figure}

\section{Conclusions}

In this paper we introduce an end-to-end embedding approach for user representation and demographic predictions based on multiple sequences describing attributes of purchases, and augmented by auxiliary relational data. Experiments on a large real-world dataset demonstrate the superiority of this method relative to standard baseline alternatives. 

Robust user representation holds promise in two additional use-cases in addition to personalization, which both relate to change in user behavior. The first, life-event detection, is an important issue in many applications where user behavior changes or is expected to change following events such as marriage. The second, detection of compromised accounts is important to security and prevention of identity theft in online accounts. Future research will focus on these domain specific applications. An additional benefit of user representation as described here is as a privacy maintaining user model, which avoids the need to use highly sensitive data such as user transaction information for the aforementioned purposes.

\bibliographystyle{ACM-Reference-Format}
\bibliography{lib} 


\begin{thebibliography}{00}


\ifx \showCODEN    \undefined \def \showCODEN     #1{\unskip}     \fi
\ifx \showDOI      \undefined \def \showDOI       #1{{\tt DOI:}\penalty0{#1}\ }
  \fi
\ifx \showISBNx    \undefined \def \showISBNx     #1{\unskip}     \fi
\ifx \showISBNxiii \undefined \def \showISBNxiii  #1{\unskip}     \fi
\ifx \showISSN     \undefined \def \showISSN      #1{\unskip}     \fi
\ifx \showLCCN     \undefined \def \showLCCN      #1{\unskip}     \fi
\ifx \shownote     \undefined \def \shownote      #1{#1}          \fi
\ifx \showarticletitle \undefined \def \showarticletitle #1{#1}   \fi
\ifx \showURL      \undefined \def \showURL       #1{#1}          \fi
\providecommand\bibfield[2]{#2}
\providecommand\bibinfo[2]{#2}
\providecommand\natexlab[1]{#1}

\bibitem[\protect\citeauthoryear{Bi, Shokouhi, Kosinski, and Graepel}{Bi
  et~al\mbox{.}}{2013}]%
        {bi2013inferring}
\bibfield{author}{\bibinfo{person}{Bin Bi}, \bibinfo{person}{Milad Shokouhi},
  \bibinfo{person}{Michal Kosinski}, {and} \bibinfo{person}{Thore Graepel}.}
  \bibinfo{year}{2013}\natexlab{}.
\newblock \showarticletitle{Inferring the demographics of search users: Social
  data meets search queries}. In \bibinfo{booktitle}{{\em Proceedings of the
  22nd international conference on World Wide Web}}. ACM,
  \bibinfo{pages}{131--140}.
\newblock


\bibitem[\protect\citeauthoryear{Culotta, Kumar, and Cutler}{Culotta
  et~al\mbox{.}}{2015}]%
        {culotta2015predicting}
\bibfield{author}{\bibinfo{person}{Aron Culotta}, \bibinfo{person}{Nirmal~Ravi
  Kumar}, {and} \bibinfo{person}{Jennifer Cutler}.}
  \bibinfo{year}{2015}\natexlab{}.
\newblock \showarticletitle{Predicting the Demographics of Twitter Users from
  Website Traffic Data.}. In \bibinfo{booktitle}{{\em AAAI}}.
  \bibinfo{pages}{72--78}.
\newblock


\bibitem[\protect\citeauthoryear{Dong, Yang, Tang, Yang, and Chawla}{Dong
  et~al\mbox{.}}{2014}]%
        {dong2014inferring}
\bibfield{author}{\bibinfo{person}{Yuxiao Dong}, \bibinfo{person}{Yang Yang},
  \bibinfo{person}{Jie Tang}, \bibinfo{person}{Yang Yang}, {and}
  \bibinfo{person}{Nitesh~V Chawla}.} \bibinfo{year}{2014}\natexlab{}.
\newblock \showarticletitle{Inferring user demographics and social strategies
  in mobile social networks}. In \bibinfo{booktitle}{{\em Proceedings of the
  20th ACM SIGKDD international conference on Knowledge discovery and data
  mining}}. ACM, \bibinfo{pages}{15--24}.
\newblock


\bibitem[\protect\citeauthoryear{Hu, Zeng, Li, Niu, and Chen}{Hu
  et~al\mbox{.}}{2007}]%
        {hu2007demographic}
\bibfield{author}{\bibinfo{person}{Jian Hu}, \bibinfo{person}{Hua-Jun Zeng},
  \bibinfo{person}{Hua Li}, \bibinfo{person}{Cheng Niu}, {and}
  \bibinfo{person}{Zheng Chen}.} \bibinfo{year}{2007}\natexlab{}.
\newblock \showarticletitle{Demographic prediction based on user's browsing
  behavior}. In \bibinfo{booktitle}{{\em Proceedings of the 16th international
  conference on World Wide Web}}. ACM, \bibinfo{pages}{151--160}.
\newblock


\bibitem[\protect\citeauthoryear{Jansen and Solomon}{Jansen and
  Solomon}{2010}]%
        {jansen2010gender}
\bibfield{author}{\bibinfo{person}{Bernard~J Jansen} {and}
  \bibinfo{person}{Lauren Solomon}.} \bibinfo{year}{2010}\natexlab{}.
\newblock \showarticletitle{Gender demographic targeting in sponsored search}.
  In \bibinfo{booktitle}{{\em Proceedings of the SIGCHI Conference on Human
  Factors in Computing Systems}}. ACM, \bibinfo{pages}{831--840}.
\newblock


\bibitem[\protect\citeauthoryear{Malmi and Weber}{Malmi and Weber}{2016}]%
        {malmi2016you}
\bibfield{author}{\bibinfo{person}{Eric Malmi} {and} \bibinfo{person}{Ingmar
  Weber}.} \bibinfo{year}{2016}\natexlab{}.
\newblock \showarticletitle{You Are What Apps You Use: Demographic Prediction
  Based on User's Apps.}. In \bibinfo{booktitle}{{\em ICWSM}}.
  \bibinfo{pages}{635--638}.
\newblock


\bibitem[\protect\citeauthoryear{Wang, Guo, Lan, Xu, and Cheng}{Wang
  et~al\mbox{.}}{2016}]%
        {wang2016your}
\bibfield{author}{\bibinfo{person}{Pengfei Wang}, \bibinfo{person}{Jiafeng
  Guo}, \bibinfo{person}{Yanyan Lan}, \bibinfo{person}{Jun Xu}, {and}
  \bibinfo{person}{Xueqi Cheng}.} \bibinfo{year}{2016}\natexlab{}.
\newblock \showarticletitle{Your cart tells you: Inferring demographic
  attributes from purchase data}. In \bibinfo{booktitle}{{\em Proceedings of
  the Ninth ACM International Conference on Web Search and Data Mining}}. ACM,
  \bibinfo{pages}{173--182}.
\newblock


\bibitem[\protect\citeauthoryear{Ying, Chang, Huang, and Tseng}{Ying
  et~al\mbox{.}}{2012}]%
        {ying2012demographic}
\bibfield{author}{\bibinfo{person}{Josh Jia-Ching Ying},
  \bibinfo{person}{Yao-Jen Chang}, \bibinfo{person}{Chi-Min Huang}, {and}
  \bibinfo{person}{Vincent~S Tseng}.} \bibinfo{year}{2012}\natexlab{}.
\newblock \showarticletitle{Demographic prediction based on users mobile
  behaviors}.
\newblock \bibinfo{journal}{{\em Mobile Data Challenge\/}}
  (\bibinfo{year}{2012}).
\newblock


\bibitem[\protect\citeauthoryear{Zhuang, Zhang, Zhang, Tantrum, Mah, Zeng,
  Chen, and Wang}{Zhuang et~al\mbox{.}}{2006}]%
        {zhuang2006demographic}
\bibfield{author}{\bibinfo{person}{Dong Zhuang}, \bibinfo{person}{Benyu Zhang},
  \bibinfo{person}{Heng Zhang}, \bibinfo{person}{Jeremy Tantrum},
  \bibinfo{person}{Teresa~B Mah}, \bibinfo{person}{Hua-Jun Zeng},
  \bibinfo{person}{Zheng Chen}, {and} \bibinfo{person}{Jian Wang}.}
  \bibinfo{year}{2006}\natexlab{}.
\newblock \bibinfo{title}{Demographic prediction using a social link network}.
\newblock   (\bibinfo{date}{Sept.~26} \bibinfo{year}{2006}).
\newblock
\newblock
\shownote{US Patent App. 11/535,160.}


\end{thebibliography}

\end{document}